# Red Cell Shape Responses to Shear Fields Evidence the Convective Organisation of 'Far From Equilibrium' Cell Membranes

J.T.Lofthouse Ph.D.

*MIRth Encryption Software plc., 21 Lytton Road, Springbourne, Dorset BH1 4SH. United Kingdom.*

EMail: *DrLofthouse@lycos.co.uk*.

**Abstract**

In a previous publication, I suggested that biological membrane lipids do not form randomised 'BiLayers'in the Far from Equilibrium state, but are spatially-organised by convective and shear-driven mechanisms. This new membrane model predicts that protein/lipid feedback would permit bifurcations in spatially-organised fluid flows to induce specific cell shape changes, relegating the cytoskeletal proteins to a *passive* role in their mechanical behaviour - a Gestalt shift in Cell Biology. **1**

In this paper, the ability of this fluid-driven mechanism to predict precisely all red cells responses to shear flows - the 'Discocyte to Elliptocyte' shape change, the 'Tank Treading' motions of particles trapped on their surface, the inverse temperature-dependency of their elastic shear modulus, and the peculiar 'Rhomboid' deformation sequence of Elliptocytic phenotypes which has puzzled Haematologists for three decades - provides further evidence of convective organisation of biological fluids, and introduces a significant dynamic dimension to studies of the growth, form and mechanics of biological systems. .

---

**1. Introduction: red cell mechanics under shear - the current paradigm**

The human red cell comprises two immiscible viscous fluid layers - the cytoplasm and membrane, and a well-characterised protein network (the 'cytoskeleton') that in the human case, forms as a predominantly hexagonal polyhedron between them. The edges of this triangulated network are composed of two antiparallel heterodimers of the proteins alpha and beta Spectrin. ( **Fig. 1A**). Each heterodimer contains one strand each of alpha and beta Spectrin, a highly flexible protein composed of multiple 'helical barrel' structures, separated by flexible 'linker' regions (**Fig. 1B**). Two of these heterodimers join at one end to network vertices (the transmembrane Junction Point Complexes: JPCs), and at the centre of each network edge via a transmembrane complex containing AE1 protein and Ankyrin.

Whilst red blood cell diameters show wide species variations (2.1 to 9.2 $\mu$m: **2,3**), all mammalian erythrocytes have the capacity to deform elastically as they negotiate the micro-circulation. Red cell deformation is traditionally studied by subjecting them to shear flows in a modified Couette-Taylor apparatus, where they are observed to undergo a Symmetry-breaking change in shape, from a biconcave disc to a prolate ellipsoid **4**(Fig. 6). Clearly, with a hexagonal cytoskeleton, this shape change demands the preferential elongation or shortening of a specific sub-set of network edges. Because the prevailing 'Standard Model' of the cell models proteins as 'Meat Meccano' system, all previous attempts to model shear deformation have sought to explain it in terms of protein conformational shifts, and/or alterations to protein-protein interactions, and have ignored the fact that the protein cytoskeleton, in isolation from its' membrane and cytoplasm, retains *none* of the mechanical or elastic properties of the intact cell. This, and the fact that cell deformability demands continuous ATP hydrolysis clearly indicate that continuous heat dissipation and the cells' fluids are involved in its' mechanical behaviour, but it is difficult to imagine how the assumed 'random' flows of cytoplasm and membrane lipid could effect a defined change in cell shape **7,8**. To Cell Biologists, directed intracellular forces can only be generated by proteins. The chief problem appears to lie in their insistence that metabolically active cells are 'Equilibrium' systems, and their lack of knowledge of the pattern-forming capabilities of 'Far from Equilibrium' viscous fluids. Speicher and Marchesi **9** for example, posited a model of cell deformation which treated network edges in the resting cell as 'random' coils that straighten when pulled, and suggested that restoration of the random configuration is " driven by the increase in Entropy associated with their random structure" (sic): thermodynamic anathemas aside, how a red cell could maintain a *defined* biconcave shape at rest, or undergo a *defined* elastic shape change under shear with flexible network edges is not explained, nor is the elongation of only a specific subset of edges. Vertessy and Steck **10** suggested cytoskeletal elasticity was dominated by *enthalpic* intra- and inter-molecular interactions, elastic behaviour deriving from the breaking of intra- and inter-molecular protein bonds during deformation and their re-formation during recovery, however, this two-dimensional model too fails to explain why only a sub-set of identical protein edges woudl be involved. More recently, McGough and Josephs **16** suggested models in which network edges are viewed as 'helical springs' that can be folded or bent by alterations in Spectrin conformation, yet posit no mechanism for effecting these. These and all other models of red cell deformation **5, 11, 12, 13, 14** implicitly treat the cell as an 'Equilibrium' system, and suggest that it is the systems' imperative to 'minimise its' Free Energy' that restores normal shape on removal of the stress. As metabolically-active cells are 'Far from Equilibrium' systems whose imperative is to minimise their rate of Entropy production, **all** are falsified by the observed temperature-dependency of the cell's Elastic Shear Modulus ($\mu$). This reflects the mechanical force needed to increase the distance between network vertices during deformation, and contrary to expectation in an Entropy-dominated system, *decreases* with increased temperature**15**.

I have suggested elsewhere that a paradigm shift is needed in Cytomechanics**1**. By presenting evidence that membrane and cytoplasmic flows are convectively organised in the metabolically-active, 'Far From Equilibrium' state, I demonstrated that continuous low-affinity interactions between the cytoskeleton and aminophospholipids would effectively 'slave' the cytoskeletal network to PS/PE flow sinks at the inner membrane surface, and would allow dissipative fluid structures to determine the contour lengths and spacing of network edges. The feedback model predicted that *bifurcations* between different convective planiforms would induce affine transformations of the cytoskeleton, and hence deform cell shape: in this paper, the ability of this new model to accurately *predict* all red cell responses to externally-applied mechanical shear is suggested to provide further evidence of Benard-Marangoni convection in the membranes of metabolically-active biological cells.

## 2. The Feedback Mechanism Of Erythrocyte Deformation

In**1**, I suggested that the lateral and 'Flip-Flop' motions of the aminophospholipids Phosphatidyl-serine (PS) and -ethanolamine (PE) in metabolically-active cell membranes evidenced their convective organisation. Since Spectrin has acyl chains added along its' length post-translationally, and undergoes low-affinity interactions with PS and PE **17, 18**., I demonstrated how protein/lipid feedback would enable spatially-organised PS/PE flows to 'crumple' the subunits of network edges together, determining their contour lengths. As shown in **Figs. 2A-C**, the planiform and wavelength of aminophospholipid flows would determine the contour length of specific sub-sets of network edges, and hence the overall shape and size of the cell. This feedback mechanism predicts that a bifurcation from a hexagonal to a roll planiform would induce an affine deformation of the cytoskeleton, and hence a defined change in cell shape.

The application of mechanical shear to convectively-organised viscous fluid layers manifesting a hexagonal planiform is known to induce a *predictable* change in flow pattern, characteristically a pitchfork bifurcation into counter-rotating rolls **22**. Roll orientation with respect to the applied shear is a function of shear *rate*: alignment being *transverse* to the direction of applied shear at low rates **21,22**, passing through a square cell intermediate to undergo a bifurcation into *parallel* rolls as shear rate increases (**Fig. 3**). As shown below, if we make a *qualitative* assumption that applying mechanical shear to a convectively-organised cell membrane can induce the same bifurcation in PS/PE flow patterns through it, discocytic red cells would deform into ellipsoids, *precisely* as observed when they are suspended in the gap of a rotating Couette-Taylor apparatus.

Consider initially a planar, triangulated section of a Spectrin network based on a T = 9 (a = 3, b = 0) lattice (**Fig. 4A**), laying beneath a section of membrane manifesting a hexagonal convective planiform (wavelength =$\lambda_1$). By virtue of low-affinity protein/lipid interactions, network edges will be dragged into upwelling lipid flows at the inner cell surface. Where convection is surface-tension driven, fluid flows in each convection cell are upwards at the centre, and at six 'nodes' running from the centre to the corners, and downwards in plumes along the perimeter. As cytoskeletal network edges are composed of anti-parallel strands of alpha and beta Spectrin, their surface topography complements the natural fluid contours at these nodes, and fluid flows can adjust the distance between individual helical subunits, 'crumpling' them together to accomodate the wavelength of the pattern (x-x = 3$\lambda$ : **Fig. 4B**).

A low rate of shear is applied to the membrane surface in the direction arrowed. Assuming this causes PS/PE convection cells to bifurcate into *transverse* counter-rotating rolls ($\lambda$ = width of two counter-rotating rolls), PS/PE flows would now drag a subset of parallel Spectrin edges towards upwellings, at a spacing equal to roll pattern $\lambda$ (circa twice the depth of the membrane; 20 nm) . No longer subject to a 'crumpling' force, the helical subunits of these edges could preferentially elongate up to a maximum contour length of 160 nm. Since Spectrin triangles are no longer equilateral, the network would distort as shown (**Fig. 4C**).

As shear rate increases further, flows will pass through a square cell intermediate until PS/PE rolls run *parallel* to the direction of applied force. Spectrin edges parallel to X-X will now be dragged into PS/PE flow sinks, and will elongate, producing the elliptic network shown in **Fig. 4D.** . Note that it is the aminophospholipid rolls that undergo the 'transverse to parallel' re-orientation relative to the direction of applied shear - Spectrins merely follow passively. No force whatsoever need be generated by the proteins of the cytoskeleton in order to deform it, as all existing models assume. If we approximate the cytoskeleton of a normal biconcave cell by 24-hedron using a T = 36 (a = 6, b = 0) lattice, the bifurcation sequence induces deformation into the ellipsoid shown in **Fig. 6**. This affine transformation is effected by the systems' fluid dynamics which act as the *primary* determinants of the shape change.

In sharp distinction from existing models, this fluid-driven mechanism accurately predicts the observed *dimensions* of deformed human erythrocytes. Cell deformability is assessed using a 'Deformability Index' (D.I.), which is operationally defined in terms of a ratio of the length of the major (x) and minor (y) axes of the deformed cell as measured from a photograph (D.I.= x - y / x + y: **24**). According to **23**, the cytoskeletal network of an intact cell has $\cong$ 100,000 (+/- 10 %) Spectrin edges and $\cong$ 35-40,000 (+/- 15 %) vertices. Using Eulers' network theory, we can therefore represent it as a 24-hedron using a T = 2704: (a = 52, b = 0 lattice) **1**. We assume that under shear:

1. Spectrin interactions with trans-membrane ligatures remain intact, and
2. Individual network edges can only elongate to a maximum contour length of 160 nm.

Based on this 24-hedral model, with Spectrins parallel to the major axis of the ellipsoid running in PS/PE roll sinks at the inner membrane surface, the *maximum* circumference of a deformed cell would be 26 x 12 x 160 = 49.92 µ m, yielding a length of 49.92/2 = 24.96 µ m as measured from a photograph. Similarly, because network edges running parallel to the major axis lie in the sinks of counter-rotating PS/PE rolls, they will be spaced at a distance equal to λ . The number of horizontal network edges around the minor axis is 52 x 6 = 312. If λ is unchanged by the bifurcation, and remains the same in the intact undeformed cell, (inter JPC distance $\cong$ 76 nm: Waugh **12**), the metric distance around the minor axis of a maximally deformed cell would be 312/2 x 76nm = 11.856 µ m, yielding a minor axis diameter of 3.773 µ m measured from a photograph. My model therefore predicts a red cell D.I. of 0.737, in excellent agreement with the data of both Sutera et al. **25**, who recorded average major and minor axis dimensions of 18.6µ m and 2.85 µ m respectively (DI = 0.73), and also with data published by Bull et al., ( DI = 0.8 **26**: NB -typographical error in decimal points on *their* **Fig. 2**).

**Fig. 5** demonstrates that cytoskeletal network symmetry contributes to mechanical behaviour, a consideration absent from *all* previous models. These polyhedral models (the first published that are consistent with Eulers' network theory), serve as a caveat to the notion that a 'normal' red cell can be defined by measuring its' deformability, and demonstrate that extensive cytoskeletal polymorphism could exist in the 'normal' population, undetected by prevailing Reductionist techniques. 'Normal' cells as currently described have a biconcave shape, $\cong$ 100,000 Spectrin tetramers/cell and $\cong$ 35-40,000 vertices/mature cell. Goldberg **27** demonstrated that several icosahedra with different symmetries can be constructed with the same number of faces (F), vertices (V) and edges (E): the same is true of networks projected onto a 24-hedron. It follows that two biconcave red cells could have an identical number of Spectrin tetramers and JPCs, yet possess disparate Spectrin lattice symmetry. Both be assessed as 'normal' using prevailing criteria, but would have different mechanical characteristics because lattice symmetry affects the deformation obtained at a given shear rate. Networks with identical F, V and E (**Figs. 5A and C**) deform to different extents under the same transformation, whilst two with different T numbers can yield the same D.I. (**Figs. 5 B and C**). Symmetry differences therefore have a profound effect on cell behaviour, but are completely overlooked in mechanical models of cell deformation (see eg Discher et al.**5**).

## 3. Cell Re-Orientation Under Shear Evidences Mechanical Coupling Between The Membrane, Cytoplasm And Environment

Cells subjected to shear are observed to re-orient relative to the vertical axis as a function of shear rate, but previous 'Statics-focussed' models have never posited any explanation for this: the 'Fluid-Dynamics' based-model presented in this paper actually *predicts* it, providing direct evidence that bifurcations in membrane flow patterns are the *primary* cause of cell shape change, and indirect evidence that membrane flows are indeed convectively organised when the cell is thermally pumped.

Red cells suspended in the gap between the cylinders of Couette-Taylor apparatus initially have random orientations to one another, but are observed to 'tumble' as the cylinder begins to rotate, aligning normal to the vertical axis of the apparatus prior to deforming into ellipsoids. Once ellipsoidal, the major axis lies normal to the vertical axis of the apparatus, but develops an *angle of inclination,* away from the horizontal, at intermediate shear rates (**Fig.7 A and B**: **4**).

In the planar system used by Graham to study the effects of shear **[22]**, the convecting medium was contained in stationary horizontal tank, and uni-directional shear was applied by sliding a glass plate across the top of the apparatus. Polygonal convection cells either formed with an orientation favouring bifurcation into rolls or did not (**Fig. 3B**). In Couette-Taylor systems such as the Ekatacytometer or Rheoscope the *direction* of shear is defined by the fluid flows within the apparatus, and changes as a function of shear rate. Qualitatively, at low rotation rates, in a cylindrical apparatus, fluid flows run ~ perpendicular to the vertical axis of the apparatus, defining a single torus. The angle of these streamlines relative to the horizontal axis increases with rotation rate, until at a critical speed (defined by the critical Taylor number), the flow bifurcates into a *series* of vortices extending up the apparatus, with opposite circulation in adjacent vortices (**Fig. 7A**). From the perspective of a red cell suspended in this apparatus therefore, both the magnitude and *direction* of the applied shear force changes as the speed of the cylinders increases.

If we assert that membrane lipid is initially convectively-organised into a hexagonal planiform, at a low shear rate, the cell will 'tumble' until convection cells are brought into an alignment favouring bifurcation into rolls running transverse to the applied shear (**Fig. 7C**). Spectrin edges running parallel to X-X will elongate at a spacing of λ , and the lattice will deform. As shear rate increases further, aminophospholipid rolls will undergo a second bifurcation to run *parallel* to the flow streamlines, and will deform the cell into an ellipsoid whose major axis runs *normal* to the cylinder axis. As shear rate increases, flow streamlines in the suspending fluid develop an angle with respect to this axis: aminophospholipid rolls in the cell membrane therefore maintain their orientation parallel to these by the entire cell physically re-orienting with respect to the vertical axis. Cell 'tumbling' in a Couette-Taylor apparatus is therefore due to changes in the flow pattern *of the fluid in which cells are suspended,* supporting the notion of mechanical coupling between the membrane and environment.

This coupling explains why several discrepancies exist in the cell deformability literature between groups who routinely use a cone-and-plate Rheoscope for their studies, and those using an Ektacytometer: the former has conical geometry, the latter cylindrical: this affects the *angle* of flow streamlines in the suspending fluids, and accounts for the differences they report in orientation behaviour of red cells under shear.

With a suspending medium viscosity of 0.047 poise, red cells do not elongate even at an applied shear stress of 300 dynes/cm $^2$, yet elongate significantly at a mere 50 dynes/cm $^2$ if the viscosity of the suspending medium is raised to 0.216 poise. Likewise,

Cell deformability emerges as function not only of externally applied shear, but importantly, of the viscosity of the cytoplasm **2,25,28,29**: at a constant external shear stress and intracellular heat dissipation rate, the *threshold* of deformation increases if cytoplasmic viscosity is increased **30**. These data indicate that biological cells *can* be usefully modelled as multi-layer convecting fluid systems, similar to those under theoretical and empirical study by others ( **31, 32** and http://glf-riker.colorado.edu/multirb.htm ). A flowing cytoplasm that is mechanically-coupled to the membrane could theoretically, exert shear on it and induce bifurcations in membrane flows that would change cell shape. The model also gives a clear indication that red cell deformation *In Vivo* will be affected by flow streamlines in the external fluid environment - the plasma. Flows within the capillaries, veins and arteries of the circulatory system are currently modelled as if these vessels are smooth walled tubes. Since these vessels are themselves lined with endothelial cells (arranged in a spiral), each of which has a surface topography, microflows at the walls are likely to be highly complex in pattern, and would change as vessels increase and decrease in diameter. This raises the possibility that red cells could be shear-deformed by the fluid flows around them as they circulate without having to make contact with the walls of capillaries as all current models assert.

## 4. A convectively-organised membrane explains the 'Tank-Treading' motion of particles on the surface of deforming cells

At a constant shear rate, red cells adopt a stable, stationary orientation in the shear field, but large particles (such as fibrin, polystyrene /latex beads or Heinz bodies) adventitiously caught on the extra- and intracellular membrane surfaces move *around* the cell surface in straight lines parallel to the major axis of the ellipsoid. The details of this motion (referred to as 'Tank Treading': **Fig. 8**) have previously been interpreted against a membrane modelled as a patternless sphere of viscous fluid, leading to descriptions of "the entire membrane rotating around the cell contents" as if the cytoplasm remains stationary during the process **4, 34, 35**. The new membrane model suggests the trajectories of markers on the surfaces of deforming red cells are defined by aminophospholipid roll flow sinks at the extracellular surface, which act as 'guide rails, defining their path across the cell surface.

The trajectories of small particles caught on the inner or outer surfaces of numerous *other* cells are also compatible with them following the sinks of convectively organised lipid flow, and with the notion of mechanical coupling between organised flow in cytoplasmic fluids and a convectively-organised membrane. Because Cell Biology presently assumes that cellular fluids are 'random', all previous models have been 'Statics-based', and have **assumed** that the only way of conveying particles in defined trajectories around the cell is for them to be transported along contractile protein elements that run against the membrane using a 'ratchet and pinion' mechanism: the new model suggests an alternative mechanism is possible. Interesting parallels to this motion can be observed during plant cell Morphogenesis. Here, protein complexes that produce Cellulose microfibrils move in defined trajectories across the cell as they extrude Cellulose fibrils: these bear and maintain a defined spatial relationship to intracellular cortical MT,laying exactly mid-way between them. The mechanism guiding their motion is presently unknown ( reviewed in Giddings and Staehelin, **36** ). Because, similar to red cell Spectrin, the cortical fibres of plant cell cytoskeletons also undergo low-affinity interactions with phospholipids, if we assert that lipid flows in metabolically-pumped plant cell membranes are similarly spatially-organised, my model provides a mechanism that would evenly space Microtubules on the inner cell surface at lipid flow sinks, *and* (since fluid trajectories in Taylor vortices are advancing spirals), could push Terminal Complexes across the cell, depositing Cellulose fibres exactly mid-way between them The feedback mechanism suggests that the cause of this classic plant cell 'Microtubule-Microfibril Syndrome' lies not in their protein components, but in the spatial organisation of their lipid flows: this is explored in detail elsewhere.

Directed particle motions on the surfaces of other cells are also compatible with membrane flows that are spatially-organised by convective and shear-driven instabilities. For example, particles of carbon adventitiously caught on the dorsal surface of metabolically active fibroblasts are observed to move away from the ruffling edge in trajectories that are either parallel to the direction of cell motion, or at an oblique angle (~20 º) to it. They come to a halt at a specific position on the cell surface, approximately over the nucleus, and from here, some move sharply at right angles, and fall off the cell **37,38**. If one allows for spatial organisation of the membrane as a viscous fluid layer, this description indicates that there is a distinct 'equator' in the flow topology of their lipid membranes, and in my opinion, renders the cell system directly comparable with spherical Couette-Taylor systems **39** .

## 5. The New Model Explains The Rhomboid Deformation Sequence of Abnormal, 'Elliptocytic' Red Cells

Erythrocytes from propositae with Hereditary Elliptocytosis (HE) are characterised by an Elliptocytic shape. Under shear in either an Ektacytometer or a Rheoscope, these undergo a peculiar 'rhomboid' deformation sequence. At *low* applied shear stress, Elliptocytes appear to rotate around their short axis, and align ~ perpendicular to the flow. As shear rate increases, they then pass through a 'square' intermediate, and finally deform into prolate ellipsoids with major axes *parallel* to the flow (**26** : **Fig. 9A**). As shear is reduced, the sequence reverses. These observations are actually *predicted* by my model, and serve as additional strong inference of spatially-organised aminophospholipid flows in biological membranes.

My model suggests red cells could maintain a stable, resting Elliptocytic morphology when metabolically-active in three ways. As shown in **Figs.9B -9D**, each would form a rhomboid intermediate under shear.

*Type 1 Elliptocytes: The Spectrin cytoskeleton can be modelled as a 24-hedron, has a predominantly hexagonal Spectrin lattice (with one of 3 symmetries), but stable PS/PE planiform in the cell membrane is a roll pattern*

As already shown (**Fig. 4**), feedback allows aminophospholipid planiform and λ to determine the contour length of network edges: cytoskeletons with a 'normal' 24-hedral geometry are deformed into Ellipsoids if PS/PE flows through the membrane above them have a stable pattern of longitudinal, counter-rotating rolls. In the case of 'normal' cells, these arise because of a shear-induced bifurcation. Theoretically therefore, a cell with a 'normal' 24-hedral cytoskeleton could have a resting Elliptocytic shape if aminophospholipids had a stable convective roll planiform through their membranes . The new model suggests Type 1 phenotypes could arise in several ways:-

1. Assuming normal JPC- and AE1/A-Spectrin ligations, and a natural lipid dynamic favouring hexagonal cells, significant deviations in PS/PE trajectories could be induced if there were a small changes in the distribution of charged residues along the edges of the Spectrin network. These could arise because of DNA sequence differences - point mutations in either α - or β -Spectrin sequences, or additions/deletions that altered the distribution of Spectrin phosphorylation sites along their length. The Type 1 Elliptocyte is compatible with several HE phenotypes recorded in the literature: Spectrin Tunis, for example, is characterised by a mutation in which a positively-charged amino acid is substituted by an uncharged residue (Arg to Trp) in the alpha I Spectrin domain ([40]). Several other Elliptocytoses are associated with loss of positively charged residues on Spectrin alpha chains: Arg residues being replaced by Ser, Leu or Cys [41, 42]. Since in this model, convective flow patterns would also be affected by the contours of the protein edges which they flow past (eg[43]), it is also compatible with an Elliptocytosis reported by LeCompte et al., [44], in which propositae have shorter beta Spectrin chains. In Spectrin Tandil Elliptocytosis, the beta chains are shortened and have lost the ability to become phosphorylated because of a deletional frameshift mutation [45]. Spectrin Tokyo Elliptocytosis (beta 220/216) has a truncated beta chain in which Ser 2060 is replaced by an Ala residue, and hence has lost a phosphorylation site [46]. Spectrin Detroit Elliptocytes have *longer* beta chains, whilst others have altered levels of alpha chains in the mature cells [47].

   Interestingly, a *reversible* Elliptocytosis of red cells is associated with lymphoblastic leukemia, chronic myeloid leukemia, and acute myeloid leukemia [48]. All these red cells have enhanced beta-Spectrin phosphorylation at diagnosis, which disappears during remission along with a return to normal morphology. This demonstration of a reversible erythrocyte membrane alteration in leukemia is entirely compatible with the feedback mechanism presented here: enhanced Spectrin phophorylation would add negative charges to the edges of the network against which PS trajects in the metabolically-active state, causing 'wobble' and bifurcations from stable hexagonal to roll trajectories, and an affine deformation of the cytoskeleton that changes cell shape.

2. Aminophospholipid in a cell could manifest a stable roll or square pattern planiform simply because the composition of their acyl chain moieties favoured it. In a cell with 'normal' Spectrin sequence and 24-hedral cytoskeletal geometry, if aminophospholipids bore different acyl chains, minute difference in the membranes' flow patterns could rendered roll/square pattern trajectories through the membrane more stable when the cell is metabolically active. The red cells from some other mammals (e.g. Llama and Camel) have a "normal" resting Elliptical shape, which appears compatible with this type of Elliptocytosis. They are known to have a higher protein to lipid ratio [3], and distinct differences in the acyl chains attached to PS and PE fractions of their lipids [96].

3. Cells might have a cytoskeleton based on a 24-hedron, but a different metabolic rate, such that the rate of internal heating favoured roll or square planiforms. Southeast Asian ovalocytes (SAO) cells, though commonly described as 'oval' in shape, share similar broken symmetries to Elliptocytic cells. They are characterised by increased ATP consumption rates, resulting in premature ATP depletion *in vitro* [49,51,50], and have an AE1 (Band 3) protein that forms abnormal strands bound to the Spectrin network. From available ultrastructural data, it is clear that these strands run parallel to one another, but it is not known whether they run latitudinally or longitudinally relative to the equator of the cell. Under the Dynamic Template model, increased rigidity is predicted in these cells, since an additional force would be required to overcome inter-Band 3 interactions in these strands before the Spectrin lattice can deform. Increased rigidity has been reported [52,53], and whilst no detailed data concerning the lipid composition of their membranes exists, Saul et al., [53] have recorded a higher micro-viscosity for Ovalocytes than for control membranes.

*Type 2 Elliptocytes: The Spectrin lattice is hexagonal (with one of 3 symmetries), but the cytoskeleton is not based on a regular 24-hedron. Stable aminophospholipid planiform through the membrane above this can be either hexagonal or roll pattern.*

Whilst some Elliptocytic cells appear to form with a normal discoid morphology, and only adopt an Elliptical shape after time in the microcirculation [54], there are some phenotypes whose cytoskeletons are reported to retain an Ellipsoidal shpae when the lipid membrane is extracted from them [55]. In the latter cases, this implies that these cells are *formed* with an abnormal geometry, and that a Symmetry-Breaking bifurcation in a pattern-forming mechanisms takes place at an earlier stage in their Development.

If based on hexagonal lattices, construction of a closed polyhedron with an Elliptocytic shape requires alteration to the number and spatial arrangement of non 6-fold vertices : **Fig. 9C** shows the top view of one possible irregular 20- hedron based on a T = 48 (a = 4: b = 4) hexagonal lattice. Stable PS/PE trajectories through the unstressed membrane above this cytoskeleton could either be roll *or* hexagonal patterns: either would support an Elliptic shape, but my model predicts that behavour under mechanical shear would differ.

Consider a simpler T = 3 ( a = 1, b = 1) lattice underneath a membrane manifesting roll pattern convection (**Fig. 9**). In a low shear field, flow streamlines around the cell would run ~horizontal to the axis of the apparatus: the entire cell would therefore initially orient so as to bring PS/PE rolls into transverse alignment across these. As shear rate increased, the *aminophospholipid rolls* would undergoes a second bifurcation, re-orientating to run parallel to the applied shear. This would bring a different set of Spectrin edges into aminophospholipid roll sinks at the inner membrane surface: these will unwind, and the cell will elongate into the direction indicated in **Fig. 9C.**

*Type 3 Elliptocytes: The cytoskeleton forms with a rectangular geometry, and has a Spectrin lattice based on a rectangular lattice: aminophospholipid trajects through the membrane above the lattice in a stable square or roll planiform.*

In some cases of Elliptocytosis, ultrastructural evidence reveals a cytoskeleton based on a *rectangular*, rather than a hexagonal lattice, ie , *4-fold* Spectrin vertices predominate [49,50]. In these cases, the stable aminophospholipid planiform in the resting, metabolically active cell would need to be either roll or square planiform. The shape deformation caused by the shear -induced re-orientation of aminophospholipid rolls in the membrane above the lattice is easier to rationalise, since Spectrin members run at right angles to each other , moreover, the lattice complements and therefore easily accommodates the 'contours' defined by the square cell fluid intermediate (Fig. 9D).

Both square **56** and rectangular convective planiforms **57** do emerge as stable solutions of the Membrane Equation. Square cell Benard Marangoni convection has also been realised empirically, using Silicon oils that have been fractionated (Schatz: personal communication - see his research pages): the feedback mechanism of cytoskeletal assembly would demands that the membrane lipids of these propositae manifest a stable square cell planiform when Spectrin is being assembled, rather than the hexagonal planiform of 'normal' cells. This would require a difference in the lipid content of the membrane, such that the viscosity/acyl chain distributions favoured a stable square planiform. There is some evidence that *the acyl chain distributions of the phospholipids* (the main determinant of viscosity) of these cells does differ from normal propositae **49, 50**, but this is insufficient at the present time to use in a quantitative model.

Elliptocytoses are, at present, grossly classified according to the degree to which cell shape deviates from discocytic (Type I-IV): no model of cytoskeletal assembly has been posited. The feedback mechanism presented in this paper suggests there are at least *ten* distinct, theoretical causes of 'Elliptocytic' phenotype, and thus, in sharp distinction from other models, suggests a reason for the often contradictory reports in the literature regarding their deformability and re-orientation behaviour under shear- different research groups use control samples from different propositae. Waugh **58**, for example, reports that Elliptocytes have a **decreased** shear rigidity, whereas Engelhardt and Sackmann **59** showed the Elliptocytes used in their studies had a two to three fold *higher* shear modulus than normal cells and a slightly higher membrane viscosity. According to Chabanel et al., **60** μ is higher for HE subjects studied, but membrane η is either higher *or* lower: the phospholipid composition of these cells remain to be examined. Elliptocytoses are associated with a *range* of protein defects, all of which may be better understood under the new model: some have point mutations in Spectrin sequences **41, 42,** 45, **55, 61**, some decreased relative abundances of JPC components 4.1, p55 and glycophorin C/D, potentially rendering JPC/Spectrin interaction weaker **62**, one has aberrant Rh status **63**, but all show an increased thermal sensitivity **64**. Whilst protein anomalies vary from one propositus to another, *all* Elliptocytoses identified to date have either qualitative differences in membrane aminophospholipid composition **46**, and/or significantly enhanced accessibility of PS and/or PE headgroups to phospholipase cleavage at the extracellular surface, indicating that dipole-dipole interactions between phospholipid headgroups, and hence PS/PE translocation patterns through the membrane differ from their 'normal' discocytic cells **1**.

Interestingly, an 'Elliptocytosis of spaceflight' has been recorded **65**: detailed here. Whilst membrane convection is, under this model, driven by surface tension, features of cytoplasmic streaming indicate a buoyancy-driven mechanism. Since the membrane and cytoplasm are clearly mechanically coupled, the latter could exert shear forces, inducing bifurcations in aminophospholipid flows, and ultimately rendering cell morphology gravity-sensitive. Some 80% of 'normal' red cells develop lipid spicules ("Stage I Echinocytes" bearing 1-5 spicules) on their surface in microgravity, but remain essentially spherical. However, in some data, a significant proportion of Elliptocytic tear drop' shaped cells have been reported. Since extensive polymorphism with regard to network symmetry, protein sequence and lipid acyl chain distribution inevitably exists within populations, and the present means of defining a 'normal' red blood cell are so gross, it would appear subtle differences are sufficient, under this model, to cause aberrant behaviour in microgravitational environment. That seemingly insignificant differences can be the source of novel behaviour in different environments may have important general implications for Evolution. A mechanism of shape change in microgravity is presented in the third paper of this series.

## 6. Transient Electric Fields Can Generate *Secondary* ElectroHydrodynamic Instabilities In Cell Membranes

The application of transient high frequency electric fields is known to exert a marked effect on red cell morphology. Depending on frequency and strength , applied fields can induce normal biconcave cells to deform into Ellipsoids **59**, or produce lipid spicules on their surface **66**, or can create transient '0.6 nm pores' in their membranes . The convective motion of electrically-conducting fluids obviously generates magnetic fields. Because aminophospholipid headgroups are negatively-charged, their convective organisation would have a similar effect.

Could organised aminophospholipid flow be caused *primarily* by the charge differential across the red cell membrane, rather than thermal convection? Available data suggests not. Removal of all Sialic acid residues from the red cell surface (and with it ≅ 90% of the negative charge on the cell surface) has no discernible influence on red cell shape, deformability or aminophospholipid 'FlipFlop' rates **67, 68** , neither does removal of haemoglobin: it thus appears that the physiologically-relevant instability causing patterned lipid flow in the membrane *In Vivo* is thermally-induced, rather than one that is primarily driven by a potential differences or fields.

Qualitatively, the application of electric fields to layers of nematics can induce roll patterns in conductive fluids, the electric force (charge x E) acting in a manner similar to the buoyancy force in thermal convection. ('Williams domains': **69** . For a sample of given thickness, the relationship between applied voltage and frequency determines a critical point (the 'dielectric relaxation frequency'), at which a transition between the Williams roll regime and a discontinuous 'chevron' pattern arises. The red cell data of Engelhardt and Sackmann **59** clearly show that the elongation of normal red cells has a parabolic relationship to both the applied field strength (maximum at ~60 kV/m) and to frequency (plateau between 500 kHz to 5 MHz). A change in red cell morphology on the application of electric fields is therefore qualitatively anticipated by the new membrane model, where these fields perturb the planform of PS/PE flow. Current 'Statics-focussed' models assume that applied fields change cell shape *exclusively* by inducing conformational transitions in proteins. Electric fields of ~$10^5$ V.cm$^{-1}$ have been shown to induce helix-coil transitions in proteins, but the new model presented in this paper strongly suggests bifurcations in aminophospholipid flow patterns contribute more to shape-change phenomena. Whilst responses to electric fields of this magnitude are of little physiological relevance in the majority of biological cells, they may be relevant when modelling shape changes in innervated tissues - muscle and neurones. It is clear that applied transient fields must perturb *both* dipole-dipole interactions between lipid headgroups (disrupting the 'surface-tension' caused by inter-headgroup bonding) and electrostatic interactions between PS and the protein network underneath. Electric fields of the order of $10^5$ V.cm$^{-1}$ can be generated at the cell surface, but with lipid headgroups oriented parallel to the membrane, corresponding dipole fields would not penetrate far into the interior. Using the 'Bi-layer' model of the membrane, Seelig et al., **70, 71, 72** calculated that re-orientation of phospholipid headgroups by about ~20 º from the surface of a bilayer would generate a dipole component perpendicular to the membrane surface of ~6 Debye, and a corresponding dipole field of $1.8.10^6$ V.cm$^{-1}$ . This is of the correct order of magnitude to initiate changes in protein conformation, but does not predict any specific change in the shape of the cell. The feedback model does.

# 7. Conclusions

This paper demonstrates how spatially-organised flows of phospholipid through the membrane ( as opposed to the 'random' trajectories currently assumed in the Fluid Mosaic model) permit fluid bifurcations to act as the *primary* agents of cell shape change. It demonstrates how the application of shear (either generated at the outer membrane surface when cells grow through tissues, or at the inner membrane surface via the flows of mechanically coupled, convectively organised cytoplasm) can alter the convective planiform of phospholipids through the membrane, deform the cytoskeleton and alter cell shape. The convectively-organised membrane model appears universally applicable: in endothelial cells subjected to fluid shear stress, cortical Actin filaments also undergo a re-alignment from 'transverse to parallel' as a function of shear rate. This observation has previously been interpreted against a model of the cytoskeleton as a protein 'Meccano' set **73**, however, the known low-affinity interaction of Actin with phospholipid **74** would permit it also to be 'slaved' to phospholipid planforms in the membrane above it, providing further evidence of a primary role for cellular fluids in effecting cell shape and its' transitions.

The model and feedback mechanism provide a superior alternative to protein based mechano-transduction models (eg **75, 76** ), as it provides a theory of cell elasticity that explains the 'anomalous' temperature-dependency of red cell deformation. Prevailing 'Entropic Spring' models of Spectrin **11, 12, 13 14** are falsified by the observed temperature dependency of the cell's elastic shear modulus (μ). This reflects the mechanical force needed to increase the contour length of network edges during cell deformation: contrary to expectation in an Entropy-dominated system, this *decreases* with increased temperature **58** .

Shape change is induced by a bifurcation in the flow pattern of aminophospholipids through the membrane. The shear force needed to induce this is consequently a function of the membrane microviscosity, and for most viscous fluids, this *decreases* with increased temperature. It follows therefore that at a higher temperature, a *lower* shear force will induce a bifurcation in aminophospholipid flow patterns through the membrane, and hence a lower force will deform the Spectrin lattice: it predicts that μ should *decrease* with increased temperature, as widely reported in the literature. The model predicts that final, maximal extension of a normal cell (determined by the maximum possible extension of circumferential Spectrin edges, and the strength of their interactions with JPC's and AE1/A) should be unchanged by increased temperature: this is found empirically. Importantly, the new model explains why raising the temperature increases the *rate* at which cells reach this maximum extension at a given applied shear (vide infra **24** ), *and* increases their measured plastic viscosity coefficient **12**. This coefficient represents the *rate* of deformation of the cell above the critical value when it starts to deform.( ie the yield shear resultant or force per unit length). At a higher temperature, a *lower* shear force would be necessary to induce the bifurcation of aminophosolipid rolls from transverse to parallel to applied shear direction, which would cause an *apparent* increase in the deformation (extension) obtained for a given applied shear.

Diverse chemical reagents are known to alter cell deformability. Prevailing 'Statics-based' models, without exception, assume these all exert their influence by affecting interactions between structural *proteins*: however, **all** these substances have a simultaneous effect on aminophospholipids, which have previously been ignored, because they consider cell membranes to have the 'random Bi-Layer' structure posited in the Fluid Mosaic model. My new feedback model suggests the rate of deformation for a given applied shear will be a function of parameters that affect the convective-organisation of the membrane lipid (ie the temperature gradient across it, membrane viscosity, viscosity of the suspending medium), and of factors affecting the surface topography of proteins. Under this model, conformational changes in protein *can* feed into the free-running dynamics of the system and affect cell 'deformability', but do so because of the effect they have on PS/PE trajectories across them. Proteins, in this cell model, act as 'corrugated surfaces' (eg. **43** ), whose contours 'force' convection, ie affect the 'Natural' convective trajectories of phospholipid.

**Glutaraldehyde** , for example, is known to increase the shear elastic modulus (μ) of normal cells by two orders of magnitude **77** . This reagent induces cross-links between amino groups in proteins, but also cross-links aminophospholipid headgroups to each other, and to Spectrin **78** , a fact neglected in previous models of visco-elasticity. Glutaraldehyde is often reported as 'an inhibitor of the aminophospholipid translocase' - but the actual experimental observation that it inhibits the trans-membrane movements of PS, which is compatible with the fluid deformation mechanism. Aged red cells (ie those that form the most dense fractions of whole blood) are markedly *less* deformable than their younger counterparts **59, 79**. During *In Vivo* lifespan, cell membranes accumulate malonyldialdehyde (MDA), a peroxidative end-product produced when reactive oxygen species [ROS] react with unsaturated lipids **80** . MDA cross-links amino groups in both Spectrin and PS headgroups, and consequently increases membrane viscosity. In the red cell, because PS and PE fractions are richer in unsaturated fatty acids than either PC or SM, they are more prone to such oxidative damage. Such cross-linking would have a dramatic effect on the ability of the lipids in 'aged' cells to 'flow', a contention that is supported by the observed alteration in aminophospholipid 'flip-flop' activity in aged human membranes **81** . Proteins (particularly Spectrin) also accumulate de-amidated residues continuously throughout lifespan, affecting both protein charge and conformation **,82, 83** , which would cause significant deviations in aminophospholipid planiform, and consequently contribute to changes in cell deformation.

**Diamide** *decreases* cell deformability, and here again, the focus of prevailing cell models is on its' ability to induce oxidative disulphide cross-bridges between protein molecules. Disturbing the hexagonal arrangement of Spectrin will obviously cause deviations in PS/PE trajectories through the membrane above them. In a previous publication **83** , I showed how 'proton transfer complexes' between aminophospholipid headgroups allow them to manifest a 'surface tension driven' convection, and how these interactions hold their headgroups 'down' in the plane of the membrane surface, occluding the sites at which phospholipases bind. Diamide also increases the amount of PS accessible to phospholipases at the extracellular surface **84, 85 , 86,** , indicating under my model, that in addition to binding Spectrin, it also causes perturbation to stable PS/PE convective motions through the membrane.

**Peroxide** treatment cross-links the protein Globin to Spectrin and other proteins **87**, but also causes significant simultaneous peroxidative damage to phospholipid, increasing membrane viscosity, and hence the membranes' ability to become convectively-organised at a given rate of metabolic pumping. Likewise, treatment of red cells with t-butyl hydroperoxide **88** or superoxide **89** which also increase micro-viscosity, also increases μ .

**Lowering pH** is reported to reduce the extension of the red cell obtained at a given applied shear rate **90** . This has previously been interpreted in terms of the effects of pH on the charge and conformation of proteins (especially Spectrin), but under the new model, one must also take into account the decrease in negative charges on PS headgroups that occurs as pH is reduced. As probed by phospholipase digestion, altering buffer pH makes more PS accessible to $PA_2$ digestion at the extracelluar surface [91], indicating that the inter-PS headgroup interactions that contribute to stable Benard-Marangoni convective motions through the membrane are perturbed by acidification.

**ATP depletion** should cause a change in cell shape under the new membrane model, because the planiform of PS/PE trajectories through the membrane, and hence cell shape, are maintained by continuous intracellular ATP hydrolysis (which generates the small but finite temperature gradient across the cell). ATP depletion invokes the 'Discocyte to Echinocyte' shape change, in which Spectrin free lipid spicules are gradually lost from the cell, but reports on the 'deformability' of Echinocytosed red cells is frequently mis-interpreted. The new membrane model does not anticipate that the *maximum possible extension* of cells should be affected by ATP depletion, since this is determined by the maximum contour length of Spectrin edges, but it does predict that for a given applied shear force, cell elongation should be **lower** if the rate of ATP hydrolysis decreases This prediction is substantiated by cells that have depleted their ATP reserves naturally **30**. The 'sphero-Echinocytes' used by Schmidt-Schonbein in their studies **35** have lost a substantial quantity of phospholipid, and therefore differ from intact cells or spiculated cells in both lipid content and dynamic, and have membranes in which the remaining aminophospholipid is not convectively-organised. The deformability of these **decreases** as a function of the extent of lipid loss **30** . Feo and Mohandas **92, 93** report that depletion of ATP with Iodoacetamide caused no change in cell 'deformability', but are referring to the maximum extension of the cell, not the rate of attaining it. Schmidt-Schonbein **35** reports that fully Echinocytosed cells (ie those that have lost lipid spicules after depleting ATP reserves metabolically), are completely undeformable. The two studies are not directly comparable because both use different methods of ATP depletion. Iodoacetamide is known to deplete intracellular glutathione, and to simultaneously cause the alkylation of erythrocyte proteins and lipids ie it causes oxidative damage to protein and lipid at the same time as ATP levels decline **94** .

'**Ageing**' is known to shift red cell deformability responses to higher electric field strengths **95** , that is, a greater field is needed to invoke the same deformation response, **59** , correlating well with the reported increase of MDA levels in membranes of aged cells, and their increased phase transition temperatures.

# 8. Acknowledgements

I wish to express my thanks to Professor Sutera for kindly providing video of Tank Treading erythrocytes.

# 9. Figures and Tables

**Fig 1: The Human Red Blood Cell Cytoskeleton**

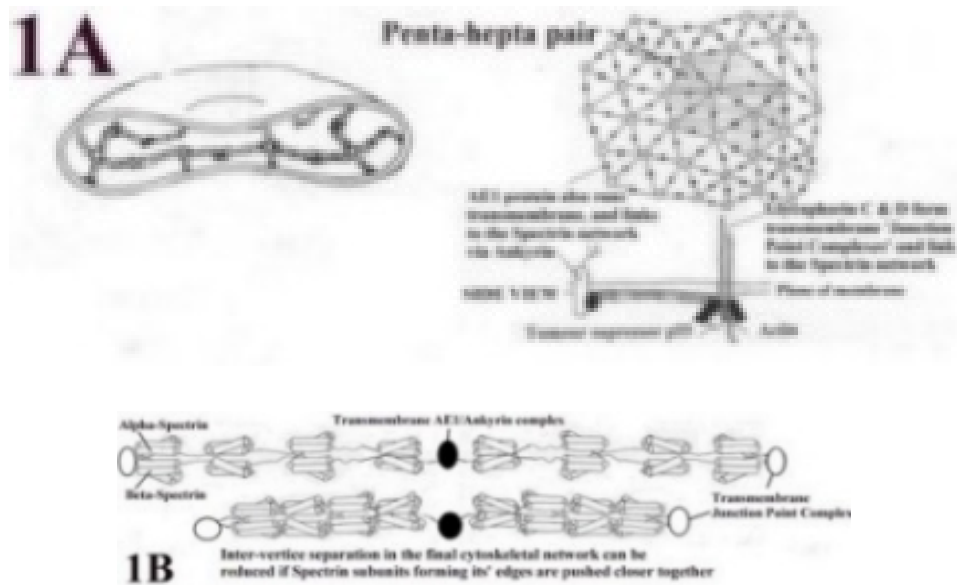

A. The red cell cytoskeleton is a network of protein that lies underneath the membrane. When extracted from the lipid membrane, this structure remains intact, retains the same shape as the cell of origin when suspended in isotonic buffer, but none of the cells' 'Tensegrity'. *In situ* , two Spectrin heterodimers join two

JPCs, with an AE1/ankyrin ligature in the centre. The predominantly hexameric network contains many 4-, 5-, and 7-fold vertices, and ubiquitous 'penta-hepta' pairs (highlighted), resembling the spontaneous defects seen in surface-tension driven fluid convection.

B. The main structural feature of both the $\alpha$- and $\beta$- chains of Spectrin heterodimers are 106-amino acid repeats (respectively 20 and 19 copies per chain), joined by flexible regions. Theoretical models predict that these fold into triple helical structures, but *In Situ* data suggests they wind around one another to form a two-start helix.

---

### Fig. 2: The Feedback Mechanism allows the contour length of cytoskeletal network edges to be adjusted by fluid flows

The flexible edges of the Spectrin network lie against the membrane, and are each composed of two Spectrin heterodimers. Edges are of equal length, but some mechanism operates *In Situ* to reduce the inter-JPC distance by a factor of 2.6 as the cell matures [1]. Prevailing models assume that the forces are generated exclusively by proteins. Since the strands of each dimer have a multi-subunit structure, spatially-organised PS/PE flows through the membrane would allow extension/contraction to be effected either by crumpling subunits closer together.

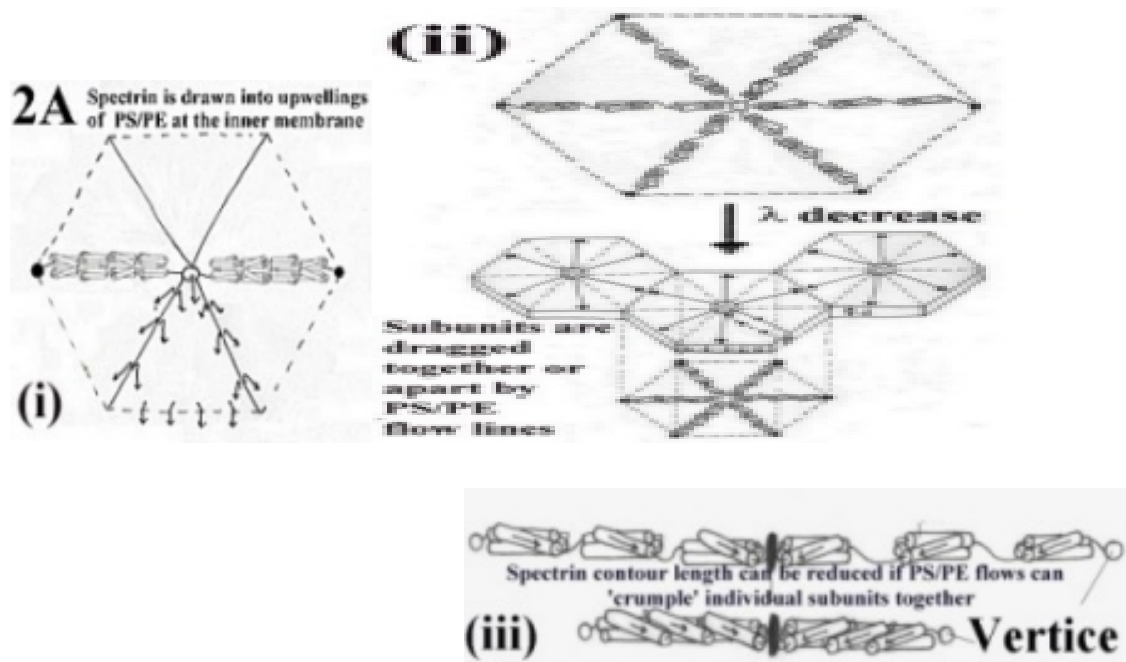

A. (i). With aminophospholipid manifesting surface-tension driven convection in a stable hexagonal planiform, network edges would lie between two flows moving in the same direction. (ii). Providing ligations to JPCs and AE1/Ankyrin ligations remain intact, the feedback model suggests changes in wavelength could be accomodated by fluid flows exerting force on subunits, moving them closer together or further apart (iii). In a wavelength expansion, all edges will be adjusted by the same amount, and shape would remain unchanged.

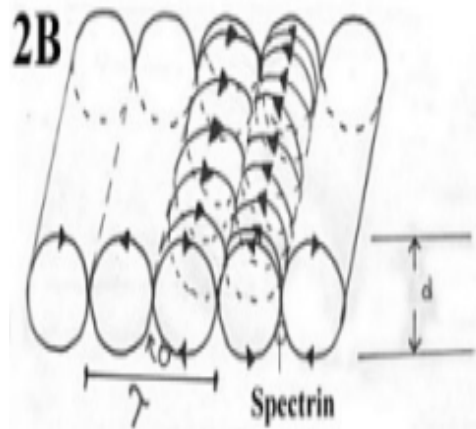

B. Counter-rotating rolls can be arise in surface-tension driven convection OR in a fluid layer subjected to mechanical shear. If laying against a membrane manifesting this planform, one set of parallel Spectrin network edges will be dragged into flow sources, and will adopt a spacing determined by PS/PE flow wavelength. Since their maximum contour length is 160 nm, whilst roll wavelengths is limited to about twice the depth of the fluid layer (d ~ 10 nm), the feedback mechanism suggests network edges would not be equal in contour length, and the lattice would be deformed.

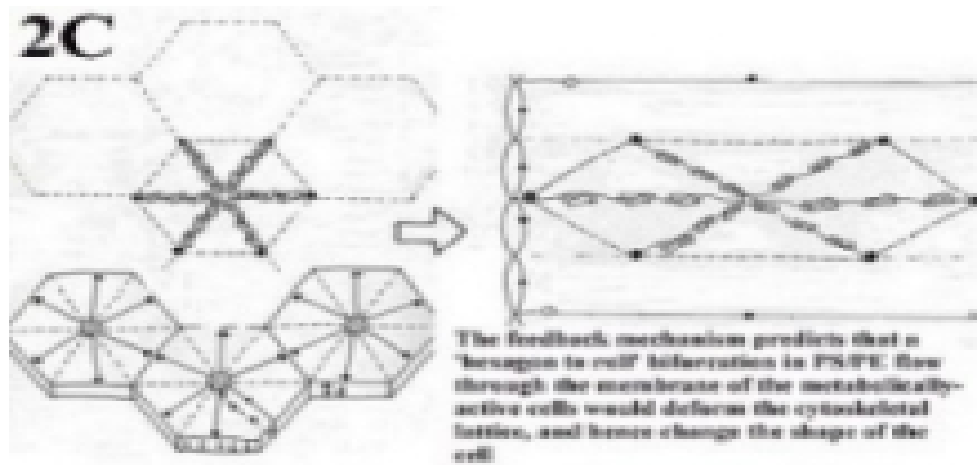

C. The feedback mechanism predicts therefore that a *bifurcation* in aminophospholipid flow patterns through the membrane above the cytoskeleton (from hexagons to rolls) would alter the relative spacing and contour lengths of individual network edges, and induce an affine deformation in the cytoskeletal lattice, and hence change the shape of the cell.

---

**Fig. 3: Effects of shear on convecting viscous media (after [22])**

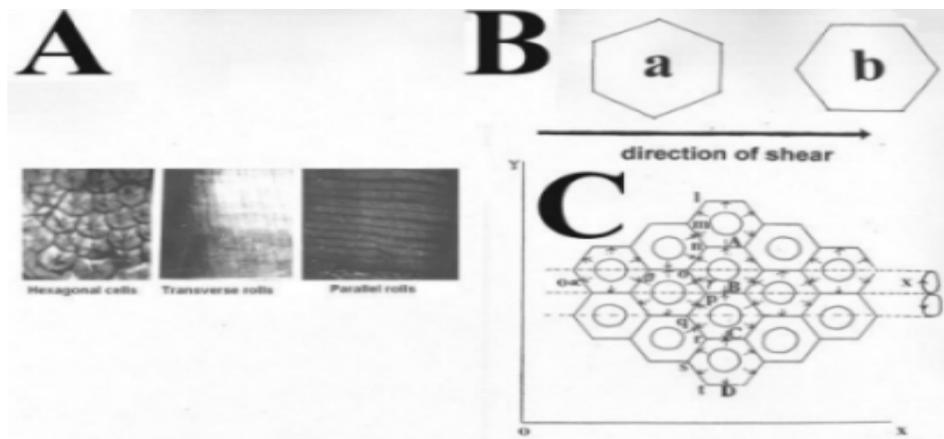

A. Graham [22] demonstrated that application of surface shear can cause a bifurcation in stable polygonal convection system, and that shear rate determines whether the induced roll pattern run transverse or parallel to the direction of applied shear.

B. Importantly, he noted also that the *orientation* of the initial hexagonal cells with respect to applied shear force determined whether longitudinal rolls formed: hexagons meeting the direction of shear in orientation (**a**) deformed, those in orientation (**b**) did not at the shear rates studied.

C. This is explained by considering a system of fluid manifesting stable hexagonal convection. Arrows represent the direction of fluid flow at the upper surface for a fluid. For shear to convert this system into counter-rotating rolls running parallel to OY, lines l – t already meet as a flow sink, and can be straightened merely by supressing flow at A, B, C, D. Conversion of hexagonal flow to a series of counter-rotating rolls parallel with OX would demand much higher shear, since adjacent sections of fluid ($\alpha\ \beta\ \chi\ \delta$ etc), are flowing in opposite directions.

---

**Fig. 4: A Pitchfork bifurcation in aminophospholipid flow trajectories an affine transformation of the cytoskeleton**

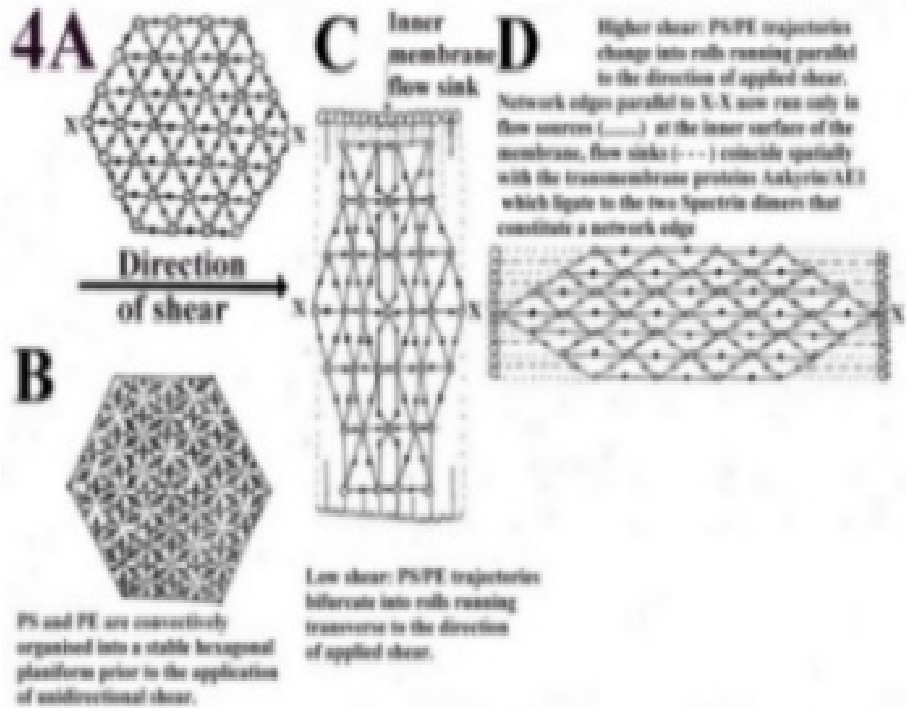

The Spectrin network based on a T = 9 (a = 3, b = 0) lattice (**A**) is 'slaved' to hexagonal aminophospholipid flow through the membrane above it (**B**). Aminophospholipids in a metabolically-active cell membrane are convectively organised into hexagonal flow trajectories above the cytoskeleton: Spectrin network edges running between adjacent cells are 'crumpled' such that their contour length equals the wavelength of the pattern. Application of low shear in the direction indicated by the arrow causes PS/PE trajectories to bifurcate into counter-rotating rolls, running transverse to fluid streamlines in the embedding fluid (**D**). Spectrins running in sources and sink are no longer 'crumpled', and are free to elongate to their maximum contour length, deforming the initially hexagonal lattice into an ellipsoid running parallel to the vertical axis of the cylinder (**C**). As shear increases further, a second bifurcation in membrane flow arises: rolls now run parallel to the direction of applied shear. This brings a different sub-set of Spectrin edges into their flow sources and sinks: these elongate, producing an ellipsoid whose major axis runs *parallel* to the direction of applied shear.

**Fig. 5: The symmetry of the cytoskeletal lattice affects its deformation under shear**

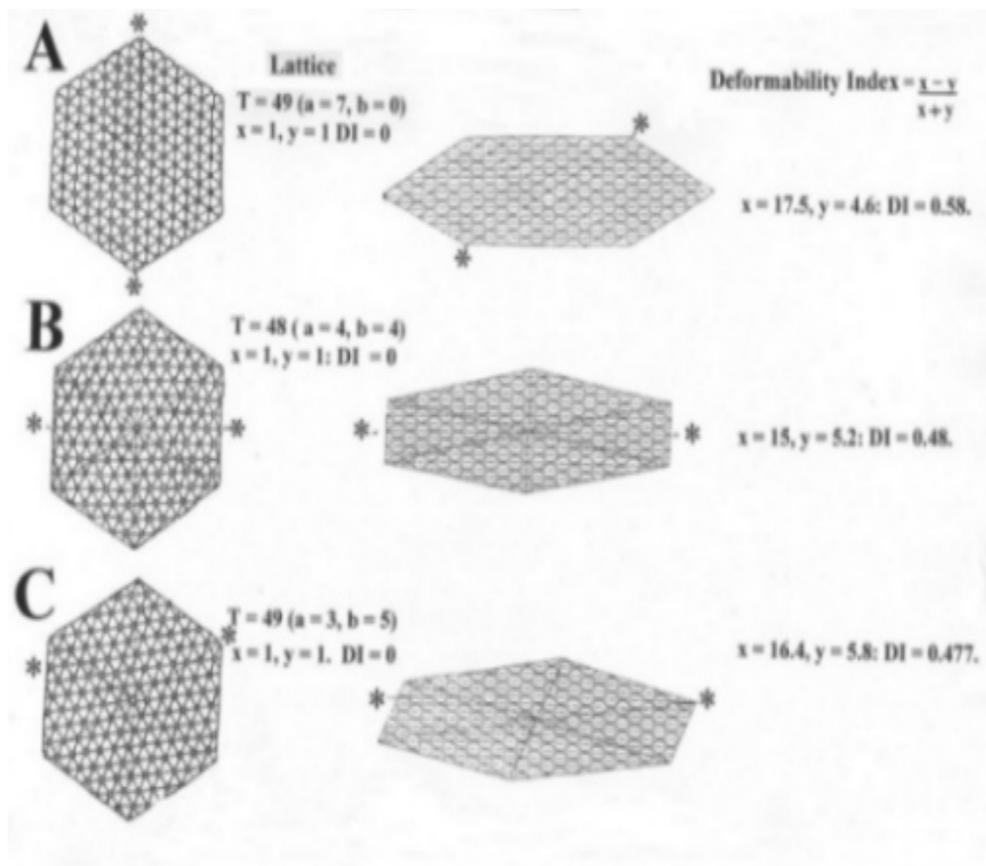

Consider the application of shear to planar hexagonal lattices based on three different T numbers. All initially lie under a membrane manifesting stable hexagonal convection with wavelength λ. Shear induces a bifurcation in PS/PE trajectories into counter-rotating rolls. Shear applied in one of three directions elongates a different subset of Spectrin dimers edges.

Current models assess deformation using a 'Deformability Index' (DI), computed by measuring the length of the major (X) and minor (Y) axes in the final ellipsoid signal (DI =X + Y / X - Y). Networks A and B both have the same number of faces, vertices and edges, yet distort under the same shear field to give differing DI values. Likewise, networks B and C deform to yield the same D.I., yet have differing T numbers and symmetries. (NB: affine transformations can take place at constant area, but area is not an affine invariant). Affine geometry highlights how misleading DI is as an indication of 'normal deformability'. The Index, as obtained using an Ektacytometer, is computed by measurement of a signal averaged from ~25,000 cells, and is in effect a ratio of averaged *diameters* of stressed and unstressed cells. As indicated by this model (Fig. 6) Spectrin network does not run across the diameter of a plane object, but *in arcs across a three-dimensional surface*. d = 2r but the length of an arc is rθ, where θ is the angle subtended at the surface in radians

**Fig. 6: Polyhedral models of a resting and deformed human red cell**

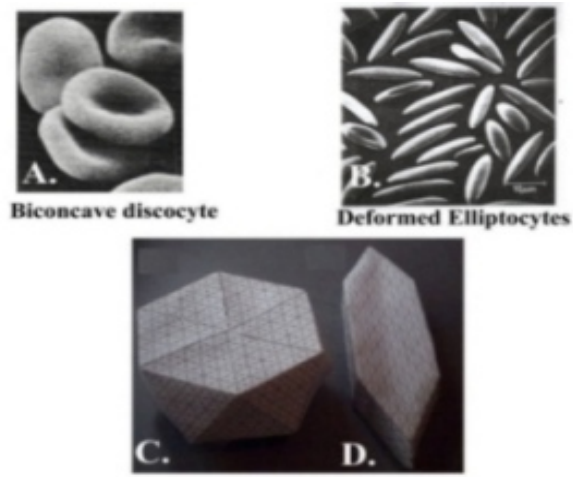

**A.** Normal biconcave red blood cell. **B.** Normal cell after shear in a Couette-Taylor apparatus. **C.** 24-hedron constructed from a T = 36 ( a = 6: b = 0) lattice. **D.** The ellipsoid generated by application of high shear rate.

---

**Fig. 7: Normocyte re-orientation as evidence of mechanical coupling with the environment**

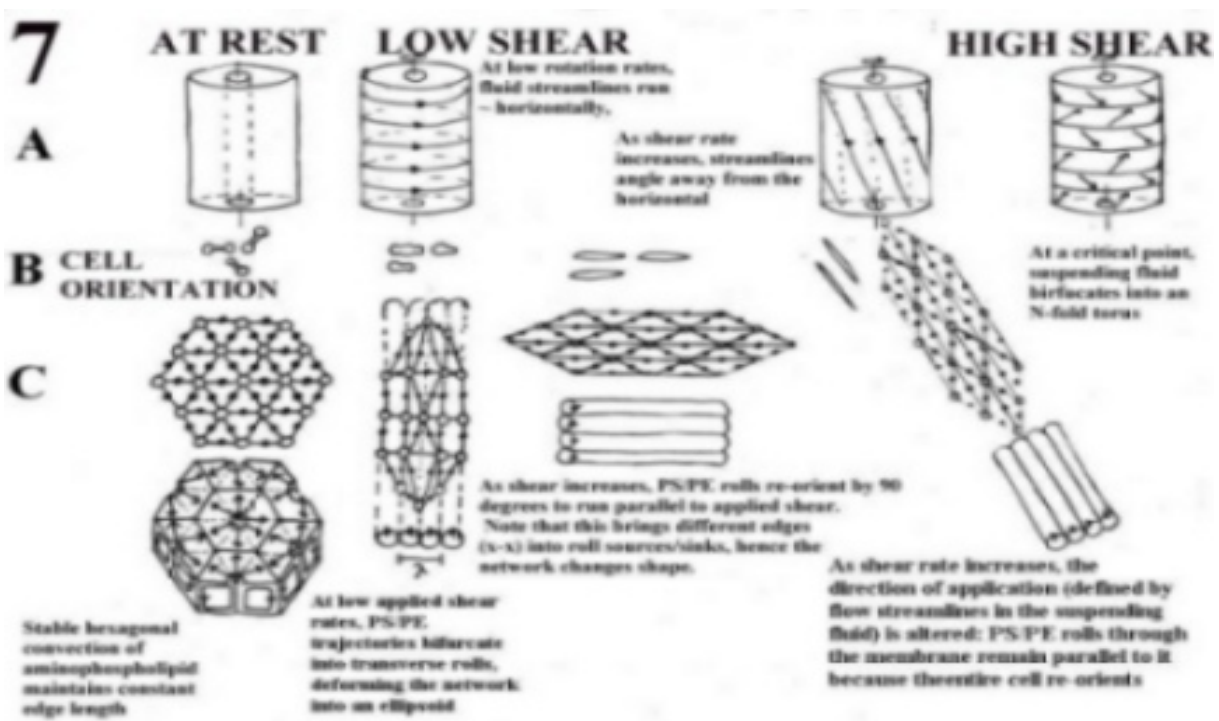

**A.** Red cell deformation is studied by suspending them between the cylinders of a Couette-Taylor apparatus. With cylinders at rest, red cells align randomly relative to the vertical axis of the system. At low rotation rates, flow streamlines are ~normal to this axis, and fluid flows in a single torus. As rotation rate increases, fluid streamlines become increasingly angled relative to this axis, until, beyond a critical point defined by the Taylor number, the fluid bifurcates into an N-fold torus.

**B.** The apparent orientation of the deformed ellipsoid changes as a function of applied shear rate.

**C.** A Spectrin network (based on a T = 9: a = 3, b = 0 lattice) is 'slaved' to aminophospholipid flow through the membrane above it (**D**). (**i**) With cylinders at rest, aminophospholipids in a metabolically-active cell membrane are convectively organised into hexagonal flow trajectories above the cytoskeleton.: network edges running between adjacent cells have their contour length determined by the wavelength of the pattern. (**ii**) Application of low shear causes PS/PE trajectories to bifurcate into counter-rotating rolls that run *transverse* to fluid streamlines in the embedding fluid. Spectrins that are forced into sources and sinks are no longer

'crumpled', and are free to elongate to their maximum contour length, deforming the hexagonal lattice into an ellipsoid running parallel to the vertical axis of the cylinder. (**iii**) As the speed of the cylinders increases, a second bifurcation in aminophospholipid flow takes place: rolls now run *parallel* to the direction of applied shear. A different set of Spectrin edges now run in the sources and sinks of aminophospholipid flows: these elongate and produce an ellipsoid with its' major axis *parallel* to the direction of applied shear. (**iv**) As shear rate increases further, the effective direction of applied shear, defined by flow streamlines in the embedding fluid changes: aminophospholipid rolls in the membrane above the Spectrin lattice maintain a parallel orientation to these, and the cell re-orients relative to the axis of the cylinder. The extension of the cytoskeletal lattice is limited by the maximum extension of network edges: above a certain shear rate, Spectrins are torn from the vertices, and the network ruptures.

---

## Fig. 8: "Tank Treading" motions demonstrate that membrane lipid flows are spatially-organised

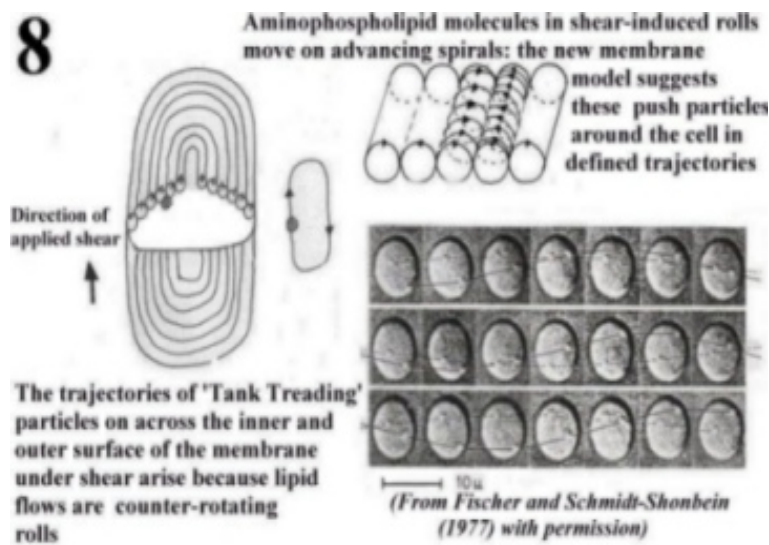

Internal and external particles adventitiously trapped against the membrane are observed to move around the surface of cells under shear in defined trajectories. Compatible with my new model, these trajectories correspond to sinks of shear induced roll patterns, and serve as further evidence of spatially organised membrane flows.

Whilst cells assume a stationary orientation in a Couette-Taylor apparatus at constant shear, the angle between the long axis of the deforming cell and the axis of the apparatus is a function of the applied shear rate. Under the new membrane model, shear induced lipid rolls running horizontal to the cell axis lie initially transverse to the vertical streamlines of the suspension medium. As shear rate increases, flow streamlines deviate from the vertical, and mechanical coupling between the fluid medium and the membrane causes the entire cell to re-orient, maintaining lipid rolls parallel to the direction of applied shear. (Tank treading photomontage from [4] with kind permission).

---

## Fig. 9: The 'Rhomboid' Deformation Of Elliptocytic Red Cells

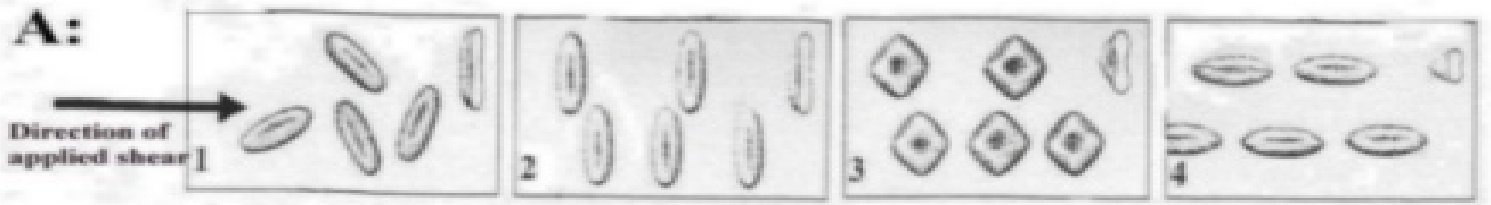

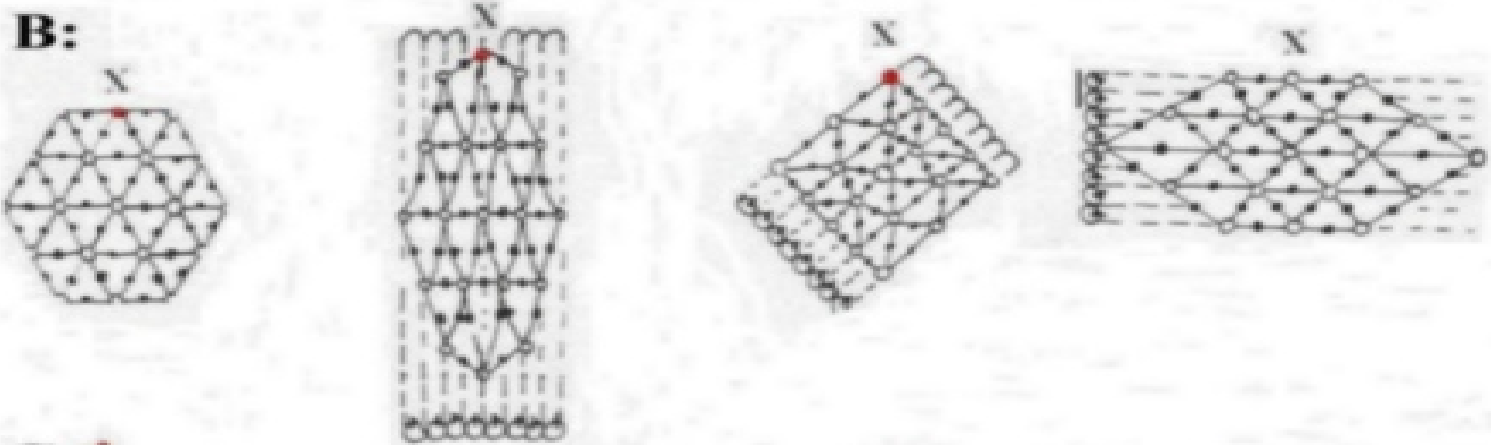

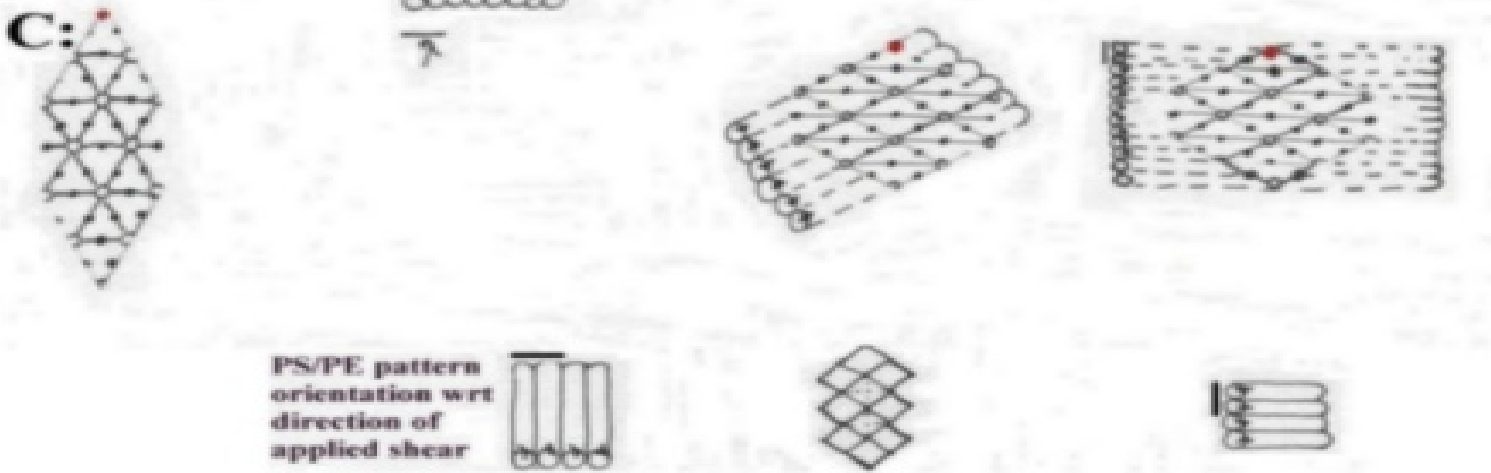

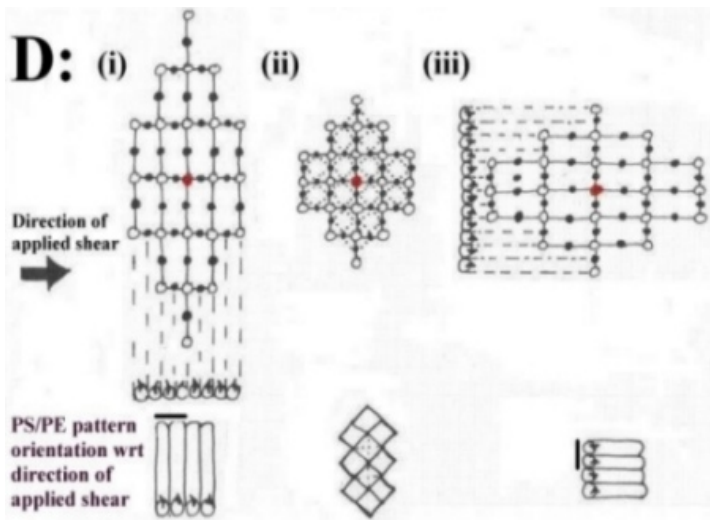

Hereditary Elliptocytosis is associated with diverse protein and phospholipid defects, but all forms are characterised by red cells that have a resting Ellipsoidal shape.

A. When placed under shear in Couette-Taylor apparatus, these first line up with their long axis ~transverse to the direction of applied shear (i), then undergo a 'rhomboid' deformation (ii), and transform into an ellipsoid with its' long axis rotated by 90 degrees, running parallel to the direction of applied shear (iii). In convecting fluid systems manifesting hexagonal planiform, application of surface shear induces a well-defined bifurcation sequence. At low shear, fluid forms counter-rotating rolls running transverse to the direction of application, passes through a square cell intermediate, then form rolls running parallel to the field. With membrane PS/PE convectively organised, feedback with the Spectrin lattice would take red cells through a rhomboid intermediate. The fact that the feedback model can mimic this puzzling red cell shape change provides even more convincing evidence that the lipid membrane **is** convectively organised when the cell is metabolically-active.

The model suggests there are several theoretical ways an intact cell could have a resting ellipsoidal shape: each would behave slightly differently under shear:-

B: The first theoretical Elliptocyte has a normal 24-hedral, hexagonal cytoskeleton, but a membrane whose lipid composition favours a convective roll planiform.

C: In the second type of Elliptocyte, Spectrin still forms a predominantly hexagonal lattice, but the overall shape of the cytoskeleton is not based on a 24-hedron. Shown here it one based on a 20 -hedron. With this type of Elliptocytosis, a symmetry-breaking event would arise at an earlier stage in Development to the third theoretical possibility:-

D: an Elliptocytic cell in which the Spectrin lattice is based on a *rectangular* lattice. The formation of a rhomboid intermediate during a 90 degree roll re-orientation is easy to visualise, since Spectrin members run at right angles to each other in the lattice shown (other rectangular symmetries are possible, and are examined in Lofthouse [1998](#)).

---